\documentclass[12pt]{article}
\pdfoutput=1
\usepackage{amssymb,amsmath}
\usepackage{graphicx}
\usepackage{color}
\usepackage[colorlinks=true
,urlcolor=blue
,citecolor=blue
,linkcolor=blue
,pagecolor=blue
,linktocpage=true
,pdfproducer=medialab
]{hyperref}
\usepackage[a4paper,width=17cm]{geometry}
\makeatletter \renewcommand{\@dotsep}{10000} \makeatother
\usepackage{appendix}
\usepackage{subcaption}
\usepackage{mwe}

\newcommand{\gmu}{\ensuremath{(g-2)_{\mu}}}

\begin{document}

\begin{center}

 {\Large  \textbf{Status Update on Selective SUSY GUT Inspired Models}
 } \vspace{1cm}

{  M. Adeel Ajaib$^{a,}$\footnote{ E-mail: adeel@udel.edu},  Ilia Gogoladze$^{b,}$\footnote{E-mail:
ilia@bartol.udel.edu\\ \hspace*{0.5cm} On  leave of absence from:
Andronikashvili Institute of Physics, 0177 Tbilisi, Georgia.},
 } \vspace{.5cm}

{ \it
$^a$ Department of Mathematics, Statistics and Physics,\\ Qatar University, Doha, Qatar \\

\vspace*{3mm}
$^b$Bartol Research Institute, Department of Physics and Astronomy, \\
University of Delaware, Newark, DE 19716, USA

} \vspace{.5cm}

\vspace{1.5cm}
 {\bf Abstract}\end{center}

We perform a status analysis of selective supersymmetric GUT models in light of recent constraints from  collider and dark matter detection experiments. We find that a significant region of the parameter space of these models is still accessible to these experiments. Amongst the models we analyze, the split family model provides solutions that can explain the observed deviation in anomalous magnetic moment of the muon. Furthermore, there is a notable region of the parameter space of each model which yields the desired relic abundance for neutralino dark matter. We also present the prediction of spin independent and spin dependent neutralino cross sections in these models and find that there is parameter space which can be probed at future experiments searching for dark matter. Our analysis serves as a motivation to continue the search for supersymmetry at various experimental fronts.

\newpage

\renewcommand{\thefootnote}{\arabic{footnote}}
\setcounter{footnote}{0}

\section{Introduction}

Supersymmetry (SUSY) is under the spotlight due to lack of evidence from extensive experimental searches at the Large Hadron Collider. Supersymmetry, however, still stands its place as one of the most intriguing theories for physics beyond the Standard Model. {Supersymmetry, particularly in the context of the Minimal Supersymmetric Standard Model (MSSM) with R-parity conservation, leads to gauge coupling unification and provides a viable candidate for cold dark matter (neutralino as the lightest supersymmetric particle (LSP)) \cite{Jungman:1995df}.} The discovery of a 125 GeV Higgs boson and the recent limits on the masses of various supersymmetric particles have resulted in severe constraints on the parameter space of SUSY.
The ATLAS and CMS experiments at 13 TeV LHC (36 fb$^{-1}$) have recently reported updated bounds on some sparticle masses. For instance,
the reported limit on the first/second generation squark masses from LHC is $m_{\tilde{q}} \simeq 1.6$ TeV \cite{lhc-squark}. The current limits on the gluino mass is $m_{\tilde{g}} \simeq 1.9$ TeV and the stop mass is $m_{\tilde{t}} \simeq 1$ TeV \cite{lhc-stop}. { In addition, current searches for the charginos have not resulted in any signals and a recent search implies a mass limit of $m_{\tilde{\chi}^\pm} \simeq 430$ GeV \cite{lhc-chargino}.} The High Luminosity LHC (HL-LHC) is expected to improve these limits if no SUSY signals are found ~\cite{upgrade,gershtein,atlaswiki,Baer:2017pba}.

With these rather severe constraints on the masses of SUSY particles and therefore its parameter space it is crucial to analyze the impact of these searches on popular SUSY models. 
{In this paper we intend to provide a status update of some selective models of SUSY by analyzing the impact of current constraints on the parameter space of these models. We focus on SUSY GUT models which will be described briefly in the following section.  Several recent studies have performed such analyses \cite{Baer:2017pba}.  }
For all the models we analyze, we find that there is still a considerable region of the parameter space that is still accessible to collider as well as direct detection experiments. In other words, it appears to be too early to give up on SUSY. We also analyze the status of dark matter in these models and find that there are regions of the parameter space of these models that yield the desired relic abundance. Another motivation of our study is the observed deviation in the muon anomalous magnetic moment
$a_{\mu}=(g-2)_{\mu}/2$ (muon $g-2$) from its SM prediction \cite{Hagiwara:2011af}
\begin{eqnarray}
\label{gg-22}
\Delta a_{\mu}\equiv a_{\mu}({\rm exp})-a_{\mu}({\rm SM})= (28.6 \pm 8.0) \times 10^{-10}.
\end{eqnarray}
In our analysis we show that the split family models allow for the $\gmu$ within the above deviation \cite{Ajaib:2014ana}.

{ The paper is organized as follows: In section \ref{sec:parameter} we  summarize the scanning procedure, the constraints we implement and the parameter space of the models we study. Section \ref{g-2} is dedicated to a brief description of
 the SUSY contribution to the muon anomalous magnetic moment  in the MSSM. In section \ref{sec:results}, we display and discuss the results of our parameter space scan. We conclude in section \ref{sec:conclude}}

\section{Phenomenological constraints and Scanning Procedure}\label{sec:parameter}

{We employ Isajet~7.84 \cite{ISAJET} interfaced with Micromegas 2.4 \cite{Belanger:2008sj} to perform random scans over the parameter space.} The function RNORMX \cite{Leva} is employed
to generate a Gaussian distribution around random points in the parameter space. We use Micromegas to calculate the relic density and $BR(b \rightarrow s \gamma)$. Further details regarding our scanning procedure can be found in \cite{Ajaib:2015ika}.  After collecting the data, we impose the following experimental constraints on the parameter space:
\begin{table}[h!]\centering
\begin{tabular}{rlc}
$123~{\rm GeV} \leq  m_h  $ & $\leq 127~{\rm GeV}$~~
\\
$ 0.8 \times 10^{-9} \leq BR(B_s \rightarrow \mu^+ \mu^-) $&$ \leq\, 6.2 \times 10^{-9} \;
 (2\sigma)$        &        \\
$2.99 \times 10^{-4} \leq BR(b \rightarrow s \gamma) $&$ \leq\, 3.87 \times 10^{-4} \;
 (2\sigma)$ &     \\
$0.15 \leq \frac{BR(B_u\rightarrow
\tau \nu_{\tau})_{\rm MSSM}}{BR(B_u\rightarrow \tau \nu_{\tau})_{\rm SM}}$&$ \leq\, 2.41 \;
(3\sigma)$ &  \\
 $ 20.6 \times 10^{-10} \leq  \Delta a_\mu  $ &$ \leq 36.6 \times 10^{-10} \,\,  (1\sigma)$ \\
 $m_{\tilde{g}}$ &$ \geq  1.9~{\rm TeV}$ \\
  $m_{\tilde{t}}$ &$ \geq  1 ~{\rm TeV}$\\
  $m_{\tilde{q}}$  &$\geq 1.6 ~{\rm TeV}$ \\
   $m_{\tilde{\chi}^\pm} $ & $ \geq  430~{\rm GeV}$.

\end{tabular}\label{table}
\end{table}

The description and ranges of the parameter space of the various models we study are as follows:

\subsubsection*{CMSSM}
The Constrained Minimal Supersymmetric Model (CMSSM) \cite{Kane:1993td}  contains the following five fundamental parameters:
\begin{eqnarray}
0 \le & m_0 &\le {\mathrm {10 \ TeV} } \nonumber \\
0 \le &M_{1/2}& \le {\mathrm {10 \ TeV} } \nonumber \\
-3 \le &A_0/m_0& \le 3 \nonumber  \\
2 \le &\tan\beta& \le 60 \nonumber \\
& \text{sign}(\mu) > 0&. \nonumber
\end{eqnarray}
{Here $m_0$ is the universal soft supersymmetry breaking (SSB) scalar mass, $m_{1/2}$ is the universal
SSB gaugino mass, and $A_0$ is the universal SSB trilinear scalar interaction (with the
corresponding Yukawa coupling factored out). The values for these three parameters are
prescribed at the GUT scale. $\tan\beta$ is the ratio of the VEVs of the two MSSM Higgs
doublets and $\mu$ is the supersymmetric bilinear Higgs parameter whose magnitude, but not the $sign$, is determined by
the radiative electroweak symmetry breaking condition. On the theoretical side, simplifying assumptions are often made for the form of the soft SUSY
breaking Lagrangian, as in the case of CMSSM inspired by minimal supergravity (mSUGRA) \cite{Chamseddine:1982jx}, which
are not fully justified on symmetry principles.
}

\subsubsection*{SU(5)}
{In the $SU(5)$ GUT, the  SM fermions of each family are allocated to the following representations: $\overline 5 \supset  (d^c, L)$ and $10\supset (Q, u^c, e^c)$, where in the brackets, we have employed standard notation for  the SM fermions.
It seems natural to consider two independent SSB scalar mass terms, at $M_{\rm G}$, $m_{\overline{5}}$ and $m_{10}$, for the matter multiplets.
The MSSM Higgs doublets belong to $5 (H_u)$ and $\overline 5 (H_{d})$ representations of $SU(5)$, which can be considered as  two independent SSB mass terms $m_{H_u}$ and $m_{H_d}$ ~\cite{Profumo:2003ema,Gogoladze:2008dk}. But here for simplicity we assume that at the GUT  scale  we have  $m_{\overline{5}} = m_{H_u} = m_{H_d}$.
Therefore, in the SU(5) scenario the  SSB masses at $M_{\rm GUT}$ are as follows:}
\begin{eqnarray}
 &&  m_{\tilde{D}^c} =  m_{\tilde{L}} = m_{H_u} = m_{H_d} = m_{\bar 5},
\nonumber \\
 &&  m_{\tilde{Q}} = m_{\tilde{U}^c} = m_{\tilde{E}^c} = m_{10} ,
\label{BC}
\end{eqnarray}
while the remaining parameters are the same as in the CMSSM. The parameter ranges for this model are:
\begin{eqnarray}
 0\leq & m_{\bar 5} & \leq 30\, \textrm{TeV} \nonumber \\
 0\leq & m_{10} & \leq 30 \, \textrm{TeV}  \nonumber \\
 0\leq & M_{1/2} & \leq 2\, \textrm{TeV} \nonumber  \\
 -3 \le & A_0/m_{\bar 5} & \le 3 \nonumber \nonumber \\
 2 \le & \tan\beta &  \le 60 \nonumber \\
&\text{sign}(\mu) > 0& \nonumber
\end{eqnarray}

\subsubsection*{NUHM2}
In the Non-Universal Higgs Model 2 (NUHM2) \cite{Ellis:2002wv}, the universality of the scalar masses is relaxed compare to the CMSSM and the Higgs SSB mass terms are assumed to be independent parameters at the GUT scale ($m_{H_u}^2 \neq m_{H_d}^2$). This parameter choice can be  realized by combining a grand unified symmetry such as SO(10) with, for instance, a
non-Abelian SU(3) flavor symmetry   acting on the three families of quarks and leptons  \cite{Babu:2014sga}. The SM Higgs field can be allocated in the 10 and 126 dimensional representation of SO(10) symmetry which naturally leads to the  non-universality in Higgs SSB mass terms. 

We consider  the following ranges of the parameters for the NUHM2 model:
\begin{eqnarray}
0 \le & m_0 &\le {\mathrm {10 \ TeV} } \nonumber \\
0 \le &M_{1/2}& \le {\mathrm {10 \ TeV} } \nonumber \\
2 \le &\tan\beta& \le 60 \nonumber \\
-3 \le &A_0/m_0& \le 3 \nonumber \\
0 \le & m_{H_u} &\le {\mathrm {10 \ TeV} } \nonumber \\
0 \le & m_{H_d} &\le {\mathrm {10 \ TeV} } \nonumber \\
&\text{sign}(\mu) > 0& \nonumber
\end{eqnarray}

\subsubsection*{NUGM + NUHM2}
For this model, in addition to employing NUHM2 boundary conditions, we assume the Non-Universal Gaugino Mass (NUGM) boundary conditions as well. {One of the elegant features of SUSY is gauge coupling unification. To retain gauge coupling unification  in the presence of non-universal   gaugino    masses at $M_{\rm GUT}$, one can employ~\cite{Martin:2009ad} non-singlet $F$-terms, compatible with the underlying GUT.  Non-universal gauginos can also be generated  from an $F$-term which is a linear combination of two distinct fields of different dimensions \cite{Martin:2013aha}.   One can also consider two distinct sources for supersymmetry breaking \cite{Anandakrishnan:2013cwa}. With many distinct possibilities available for realizing non-universal gaugino masses   while keeping universal sfermion mass at $M_{\rm GUT}$, we employ three independent masses for the  MSSM  gauginos in our study. Note that in the framework of NUGM + NUHM2, the little hierarchy problem can be easily  resolved \cite{Gogoladze:2012yf},  but we will not address this realization of natural susy in this paper.}
 Following are the ranges of the parameters for NUGM + NUHM2 model:
\begin{eqnarray}
0 \le & m_0 &\le {\mathrm {10 \ TeV} } \nonumber \\
2 \le &\tan\beta& \le 60 \nonumber \\
-3 \le &A_0/m_0& \le 3 \nonumber \\
0 \le & m_{H_u} &\le {\mathrm {3 \ TeV} } \nonumber \\
0 \le & m_{H_d} &\le {\mathrm {3 \ TeV} } \nonumber \\
0 \le & M_1 &\le {\mathrm {3 \ TeV} } \nonumber \\
0 \le & M_2 &\le {\mathrm {3 \ TeV} } \nonumber \\
0 \le & M_3 &\le {\mathrm {3 \ TeV} } \nonumber  \\
&\text{sign}(\mu) > 0& \nonumber
\end{eqnarray}
Here   $M_{1}$, $M_{2}$, and $M_{3}$ denote the SSB gaugino masses for $U(1)_{Y}$, $SU(2)_{L}$ and $SU(3)_{c}$ respectively.

\subsubsection*{SPLIT FAMILY WITH NUHM2 (SF + NUHM2)}
{Recently, Ref. \cite{Babu:2014sga, Babu:2014lwa} proposed a class of supersymmetric models in the framework of gravity mediated supersymmetry breaking \cite{Chamseddine:1982jx},  in which symmetry considerations alone dictate the form of the SSB mass terms for the sfermions \cite{Babu:2014sga,Babu:2014lwa}.  It is called flavor symmetry based MSSM (sMSSM), and in this framework the first two family sfermion masses have degenerate masses at $M_{{\rm GUT}}$, while the SSB mass term for third family sfermions is different. It was shown in \cite{Baer:2004xx} that constraints from flavor changing neutral current (FCNC) processes, for the case when third
generation sfermion masses are split from masses of the first and second generations,
are very mild and easily satisfied. This approach therefore allows for significantly lighter first two family sfermions, while keeping the third generation sfermions relatively heavy. In  SF+NUHM2 framework we consider the following parameter range for SSB terms:}
\begin{eqnarray}
0 \le & m_{16_{1,2}} &\le {\mathrm {30 \ TeV} } \nonumber \\
0 \le & m_{16_{3}} &\le {\mathrm {30 \ TeV} } \nonumber \\
0 \le & M_{1/2} &\le 2 {\mathrm {\ TeV} } \nonumber \\
3 \le &\tan\beta& \le 55 \nonumber \\
-3 \le &A_0/m_{16_{3}} & \le 3 \nonumber \\
0 \le & m_{H_u} &\le {\mathrm {30 \ TeV} } \nonumber \\
0 \le & m_{H_d} &\le {\mathrm {30 \ TeV} } \nonumber \\
&\text{sign}(\mu) > 0. & \nonumber
\end{eqnarray}
Here $ m_{16_{1,2}}$  is the common  SSB mass term for the sfermions of the first and second families, whereas the third generation sfermions have a universal SSB mass term $ m_{16_{3}}$.

\subsubsection*{SPLIT FAMILY WITH NUGM (SF + NUGM)}
{Here we consider a combination of the scenarios described above, namely,  NUGM + NUHM2 and SF + NUHM2.}  The parameter ranges are as follows:
\begin{eqnarray}
0 \le & m_{16_{1,2}} &\le {\mathrm {1 \ TeV} } \nonumber \\
0 \le & m_{16_{3}} &\le {\mathrm {5 \ TeV} } \nonumber \\
-1 \le & M_1 &\le 0 {\mathrm {\ TeV} } \nonumber \\
-1 \le & M_2 &\le 0 {\mathrm {\ TeV} } \nonumber \\
0 \le & M_3 &\le {\mathrm {5 \ TeV} } \nonumber \\
3 \le &\tan\beta& \le 55 \nonumber \\
-3 \le &A_0/m_{16_{3}} & \le 3 \nonumber \\
0 \le & m_{10} &\le {\mathrm {5 \ TeV} } \nonumber \\
&\text{sign}(\mu) < 0& \nonumber
\end{eqnarray}
For all the models above we choose the central value $m_t = 173.1\, {\rm GeV}$ \cite{:2009ec}.

\section{\label{g-2}The Muon Anomalous Magnetic Moment}

The leading  contribution from low scale supersymmetry  to the muon anomalous magnetic moment is given by \cite{Moroi:1995yh, Martin:2001st}:

\begin{eqnarray}
\label{eq:gm2}
\Delta a_\mu &=& \frac{\alpha \, m^2_\mu \, \mu\,  \tan\beta}{4\pi} {\bigg \{ }
\frac{M_{2}}{ \sin^2\theta_W \, m_{\tilde{\mu}_{L}}^2}
\left[ \frac{f_{\chi}(M_{2}^2/m_{\tilde{\mu}_{L}}^2)-f_{\chi}(\mu^2/m_{\tilde{\mu}_{L}}^2)}{M_2^2-\mu^2} \right]
\nonumber\\
&+&
\frac{M_{1} }{ \cos^2\theta_W \, (m_{\tilde{\mu}_{R}}^2 - m_{\tilde{\mu}_{L}}^2)}
\left[\frac{f_{N}(M^2_1/m_{\tilde{\mu}_{R}}^2)}{m_{\tilde{\mu}_{R}}^2} - \frac{f_{N}(M^2_1/m_{\tilde{\mu}_{L}}^2)}{m_{\tilde{\mu}_{L}}^2}\right] \, {\bigg \} },
\end{eqnarray}
 where $\alpha$ is the fine-structure constant, $m_\mu$ is the muon mass, $\mu$ denotes  the bilinear Higgs mixing term, and $\tan\beta$ is the ratio of the vacuum expectation values (VEV) of the MSSM Higgs doublets. $M_1$ and $M_2$ denote the $U(1)_Y$ and $SU(2)$ gaugino masses respectively, $\theta_W$  is the weak mixing angle, and $m_{\tilde{\mu}_{L}}$ and $m_{\tilde{\mu}_{R}}$ are the left and right handed smuon masses. The loop functions are defined as follows:
\begin{eqnarray}
f_{\chi}(x) &=& \frac{x^2 - 4x + 3 + 2\ln x}{(1-x)^3}~,\qquad ~f_{\chi}(1)=-2/3, \\
f_{N}(x) &=& \frac{ x^2 -1- 2x\ln x}{(1-x)^3}\,,\qquad\qquad f_{N}(1) = -1/3 \, .
\label{eq:gm2b}
\end{eqnarray}

The first term in equation (\ref{eq:gm2}) stands for the dominant contribution coming from one loop diagram with charginos (Higgsinos and Winos), while the second term describes inputs from bino-smuon loop.

\section{Results and Analysis}\label{sec:results}

In this section, we present our results for the parameter space scan of various models described in section \ref{sec:parameter}.
\textit{Gray} points in all the figures are consistent with radiative electroweak symmetry breaking (REWSB) and neutralino LSP. \textit{Green} points form a subset of the \textit{gray} points and satisfy the sparticle mass and B-physics constraints described in section \ref{sec:parameter}. \textit{Brown} points form a subset of the \textit{green} points and satisfy $0.001 \le \Omega h^2 \le 1 $. We choose a wider range for the relic density due to the uncertainties involved in the numerical calculations of various spectrum calculators. Moreover, dedicated scans within the \textit{brown} regions can always yield points compatible with the current WMAP range for relic abundance.
\textit{Orange} points are subset of the green points and satisfy the muon $g-2$ constraint described in section  \ref{sec:parameter}.

\begin{figure}
 \begin{subfigure}{.49\textwidth}
        \centering
        CMSSM
        \includegraphics[width=7.5cm, height=5.5cm]{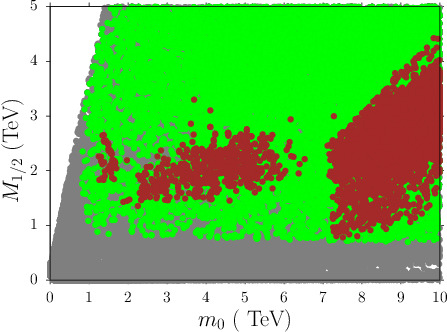}
    \end{subfigure}\hfill
\vspace{.75cm}
    \begin{subfigure}{.49\textwidth}
        \centering
       NUHM2
        \includegraphics[width=7.5cm, height=5.5cm]{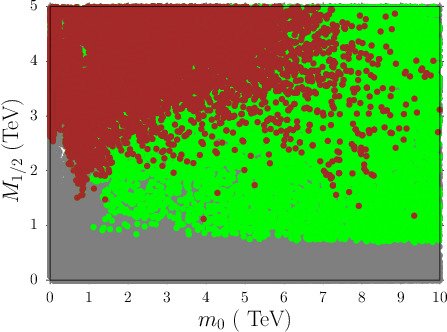}
    \end{subfigure}
\vspace{.75cm}
        \begin{subfigure}{.49\textwidth}
        \centering
       SU(5)
        \includegraphics[width=7.5cm, height=5.5cm]{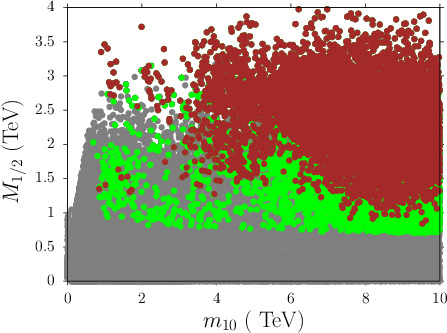}
    \end{subfigure}
        \begin{subfigure}{.49\textwidth}
        \centering
       SPLIT FAMILY NUHM2
        \includegraphics[width=7.5cm, height=5.5cm]{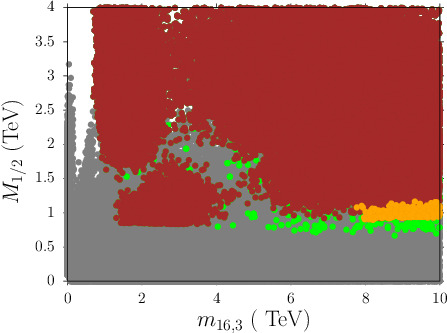}
    \end{subfigure}
    \begin{subfigure}{.49\textwidth}
        \centering
       NUGM+NUHM2
        \includegraphics[width=7.5cm, height=5.5cm]{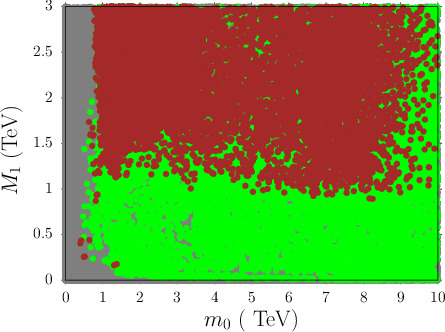}
    \end{subfigure}
        \begin{subfigure}{.49\textwidth}
        \centering
       SPLIT FAMILY NUGM
        \includegraphics[width=7.5cm, height=5.5cm]{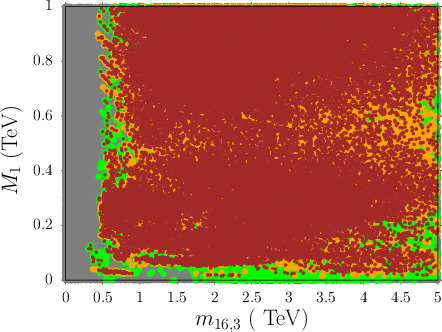}
    \end{subfigure}
\caption{Plots in the $M_{1/2}$ vs. $m_0$ plane for the models described in Section \ref{sec:parameter}. \textit{Gray} points in all the figures represent raw data. \textit{Green} points form a subset of the \textit{gray} points and satisfy the sparticle mass constraints and B-physics constraints described in Section \ref{sec:parameter}. \textit{Brown} points form a subset of the \textit{green} points and satisfy $0.001 \le \Omega h^2 \le 1 $. \textit{Orange} points are subset of the green points and satisfy the muon $g-2$ constraint described in Section  \ref{sec:parameter}.}
\label{fig:mhf-m0}
\end{figure}

\begin{figure}
 \begin{subfigure}{.49\textwidth}
        \centering
        CMSSM
        \includegraphics[width=\linewidth]{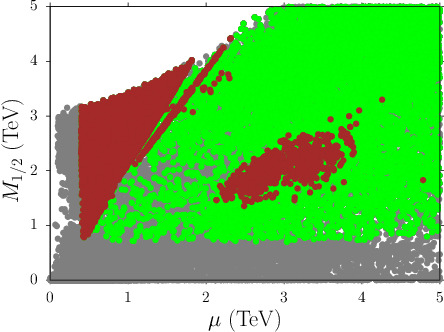}
    \end{subfigure}\hfill
\vspace{1cm}
    \begin{subfigure}{.49\textwidth}
        \centering
       NUHM2
        \includegraphics[width=\linewidth]{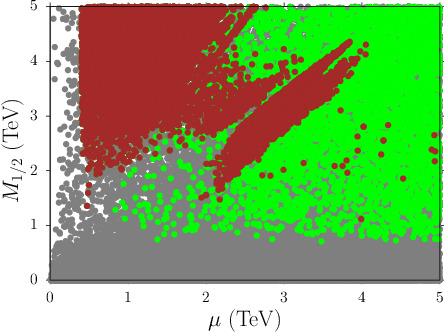}
    \end{subfigure}
\vspace{1cm}
        \begin{subfigure}{.49\textwidth}
        \centering
       SU(5)
        \includegraphics[width=\linewidth]{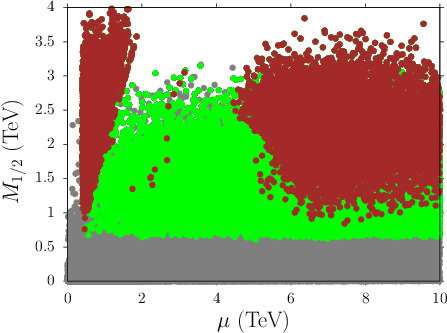}
    \end{subfigure}
        \begin{subfigure}{.49\textwidth}
        \centering
       SPLIT FAMILY NUHM2
        \includegraphics[width=\linewidth]{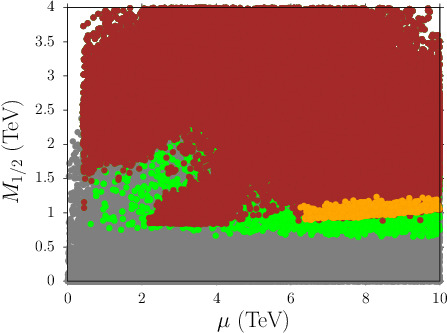}
    \end{subfigure}
    \begin{subfigure}{.49\textwidth}
        \centering
       NUGM+NUHM2
        \includegraphics[width=\linewidth]{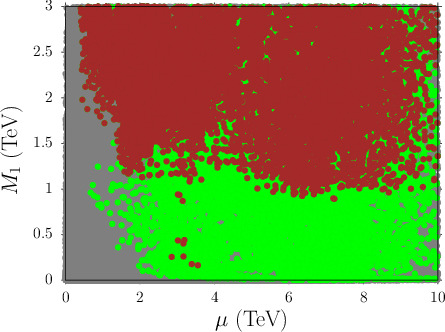}
    \end{subfigure}
        \begin{subfigure}{.49\textwidth}
        \centering
       SPLIT FAMILY NUGM
        \includegraphics[width=\linewidth]{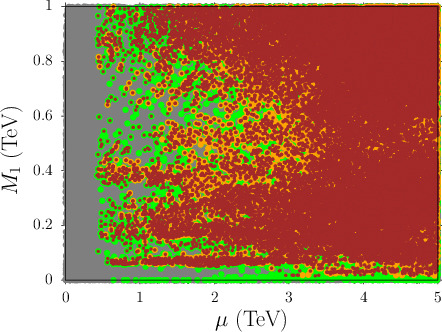}
    \end{subfigure}
\caption{Plots in the $M_{1/2}$ vs. $\mu$ plane. Color coding is the same as in Figure \ref{fig:mhf-m0}.}
\label{fig:mhf-mu}
\end{figure}

\begin{figure}
 \begin{subfigure}{.49\textwidth}
        \centering
        CMSSM
        \includegraphics[width=\linewidth]{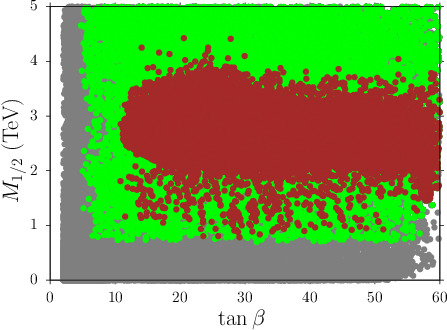}
    \end{subfigure}\hfill
\vspace{1cm}
    \begin{subfigure}{.49\textwidth}
        \centering
       NUHM2
        \includegraphics[width=\linewidth]{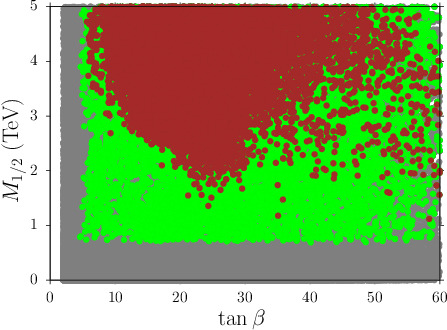}
    \end{subfigure}
\vspace{1cm}
        \begin{subfigure}{.49\textwidth}
        \centering
       SU(5)
        \includegraphics[width=\linewidth]{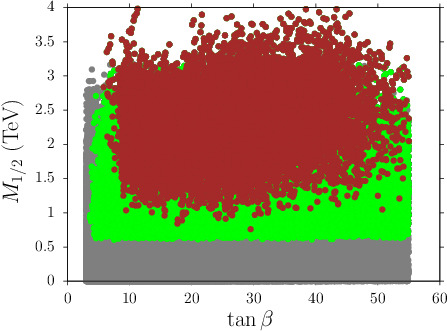}
    \end{subfigure}
        \begin{subfigure}{.49\textwidth}
        \centering
       SPLIT FAMILY NUHM2
        \includegraphics[width=\linewidth]{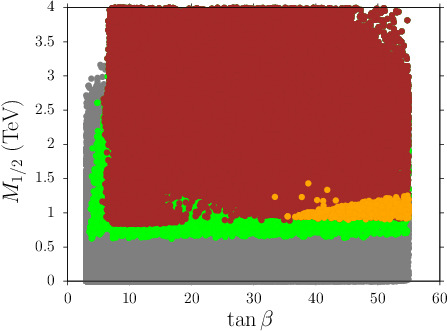}
    \end{subfigure}
    \begin{subfigure}{.49\textwidth}
        \centering
       NUGM+NUHM2
        \includegraphics[width=\linewidth]{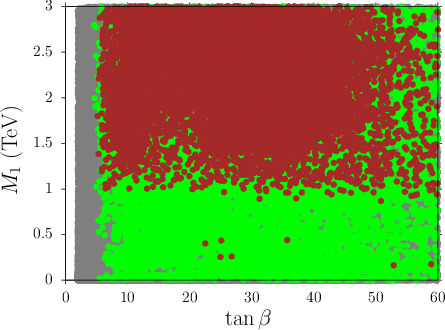}
    \end{subfigure}
        \begin{subfigure}{.49\textwidth}
        \centering
       SPLIT FAMILY NUGM
        \includegraphics[width=\linewidth]{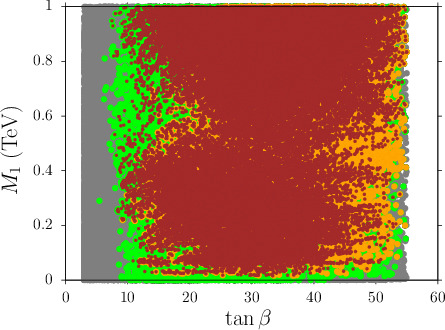}
    \end{subfigure}
\caption{Plots in the $M_{1/2}$ vs. $\tan\beta$ plane. Color coding is the same as in Figure \ref{fig:mhf-m0}. }
\label{fig:mhf-tanb}
\end{figure}

In Figures \ref{fig:mhf-m0}, \ref{fig:mhf-mu} and \ref{fig:mhf-tanb}, we plot $M_{1/2}$ vs. $m_0$, $M_{1/2}$ vs. $\mu$ and $M_{1/2}$ vs. $\tan\beta$ planes. The plots display subset of the ranges described in section \ref{sec:parameter}. The color coding of the points is as described above. We can observe that there is a considerable region of the parameter space that satisfies the sparticle mass and B-physics constraints. In addition, there is a notable region of the parameter space that yields the desired dark matter relic abundance (\textit{green} points). The $g-2$ constraint (\textit{orange} points) is only satisfied for the split family models. Except for the NUGM and split family NUGM models, the lower bound on the gaugino mass parameter is $M_{1/2} \gtrsim 1 $ TeV. It can also be noted that there is a lower bound on the third generation scalar masses, $m_{16,3}\gtrsim 8$ TeV, for the SF+NUHM2 model if we impose the $g-2$ constraint (\textit{orange} points).

\begin{figure}
 \begin{subfigure}{.49\textwidth}
        \centering
        CMSSM
        \includegraphics[width=\linewidth]{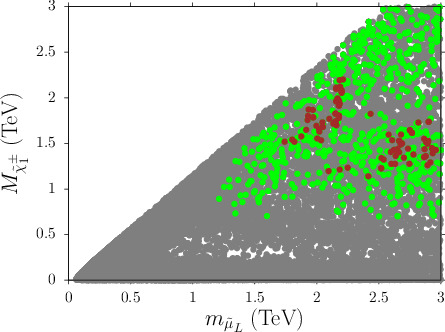}
    \end{subfigure}\hfill
\vspace{1cm}
    \begin{subfigure}{.49\textwidth}
        \centering
       NUHM2
        \includegraphics[width=\linewidth]{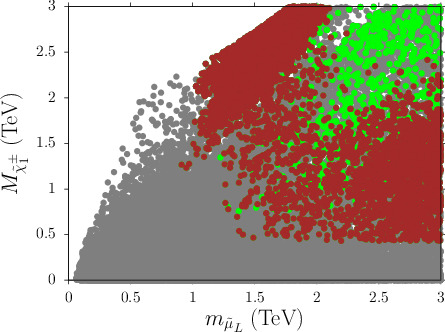}
    \end{subfigure}
\vspace{1cm}
        \begin{subfigure}{.49\textwidth}
        \centering
       SU(5)
        \includegraphics[width=\linewidth]{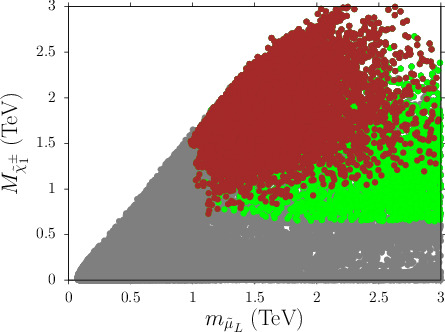}
    \end{subfigure}
        \begin{subfigure}{.49\textwidth}
        \centering
       SPLIT FAMILY NUHM2
        \includegraphics[width=\linewidth]{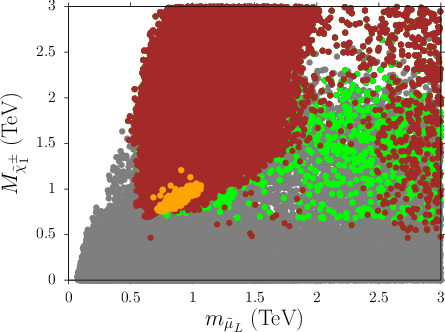}
    \end{subfigure}
    \begin{subfigure}{.49\textwidth}
        \centering
       NUGM+NUHM2
        \includegraphics[width=\linewidth]{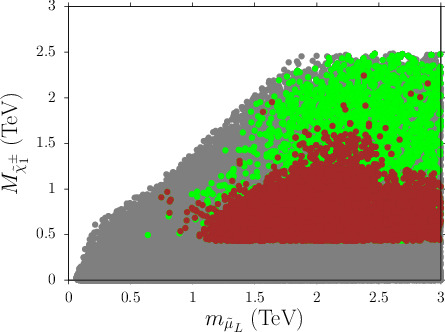}
    \end{subfigure}
        \begin{subfigure}{.49\textwidth}
        \centering
       SPLIT FAMILY NUGM
        \includegraphics[width=\linewidth]{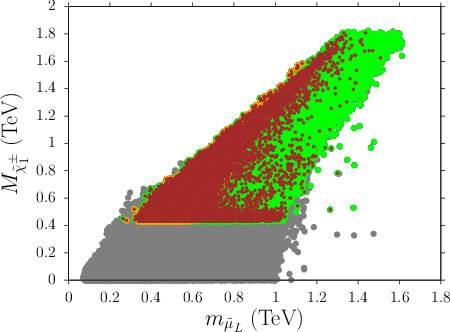}
    \end{subfigure}
\caption{Plots in the $M_{\tilde{\chi}_1^\pm}$ vs. $m_{\tilde{\mu}_L}$ plane. Color coding is the same as in Figure \ref{fig:mhf-m0}.}
\label{fig:mch-msmu}
\end{figure}

In our analysis we find that there is a significant region of the parameter space of the SF+NUGM model that satisfies the $\gmu$ constraint (\textit{orange} points). We also found good $\gmu$ solutions in a narrow region of the parameter space of the SF+NUHM2 model.  It has been found in earlier studies that there are several factors that can lead to a large SUSY contribution to $\gmu$. These include the case when $M_1$, $M_2$ and $\mu$ have the same sign~\cite{Pokoroski}, in which case both of the terms arising from chargino-sneutrino and bino-smuon loops in equation (\ref{eq:gm2}) will be positive. Furthermore, the smuons have to be light in order to make a sizable contribution to $\gmu$. For both of these models, as can be seen from Fig \ref{fig:mch-msmu}, the smuon $\tilde{\mu}_L$ is light. Note that, among other factors discussed above, large values of $\mu\tan\beta$ can lead to large $\Delta a_\mu$  as can be seen from equation (\ref{eq:gm2}). This can be seen for the SF+NUHM2 models where  $\mu \gtrsim 6$ TeV and $\tan\beta \gtrsim 30$.

\begin{figure}
 \begin{subfigure}{.49\textwidth}
        \centering
        CMSSM
        \includegraphics[width=\linewidth]{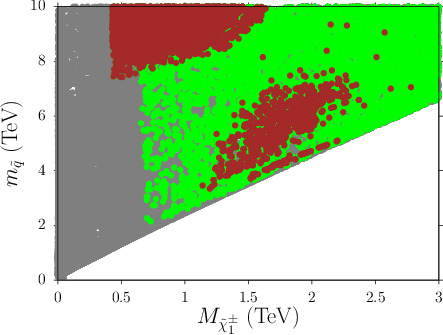}
    \end{subfigure}\hfill
\vspace{1cm}
    \begin{subfigure}{.49\textwidth}
        \centering
       NUHM2
        \includegraphics[width=\linewidth]{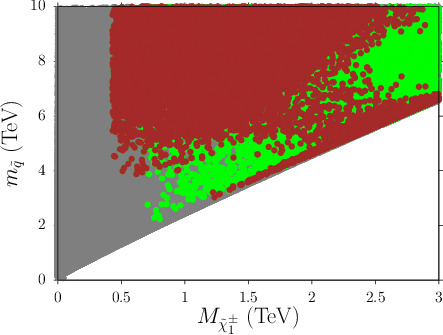}
    \end{subfigure}
\vspace{1cm}
        \begin{subfigure}{.49\textwidth}
        \centering
       SU(5)
        \includegraphics[width=\linewidth]{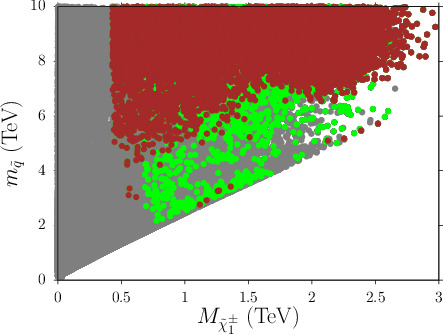}
    \end{subfigure}
        \begin{subfigure}{.49\textwidth}
        \centering
       SPLIT FAMILY NUHM2
        \includegraphics[width=\linewidth]{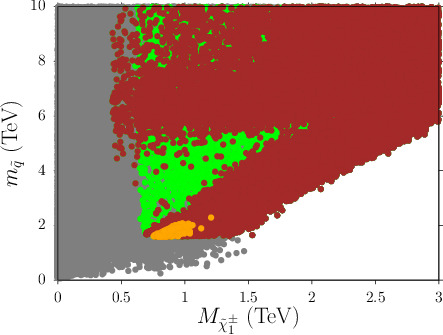}
    \end{subfigure}
    \begin{subfigure}{.49\textwidth}
        \centering
       NUGM+NUHM2
        \includegraphics[width=\linewidth]{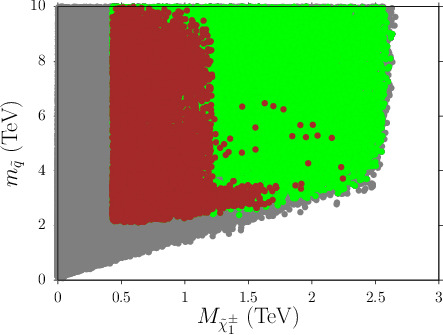}
    \end{subfigure}
        \begin{subfigure}{.49\textwidth}
        \centering
       SPLIT FAMILY NUGM
        \includegraphics[width=\linewidth]{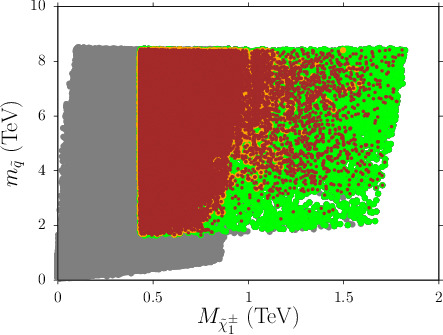}
    \end{subfigure}
\caption{Plots in the $m_{\tilde{q}}$ vs. $M_{{\widetilde{\chi}}_1^{\pm}}$ plane. Color coding is the same as in Figure \ref{fig:mhf-m0}.}
\label{fig:msq-mx}
\end{figure}

In Figure \ref{fig:msq-mx} we display our results in the $m_{\tilde{q}}$ vs. $M_{{\widetilde{\chi}}_1^{\pm}}$ plane. We can observe that, except for the SF+NUHM2 model, the lower bound on the first/second generation squark mass is $\sim 2$ TeV. The SF+NUHM2 model yields solutions with light squarks which satisfy the relic density and the $g-2$ constraint. For the SF+NUHM2 model the $\gmu$ constraint (\textit{orange} points) prefer relatively light squark masses with an upper bound of $\sim$ 2 TeV. We can see that there is a considerable region of the parameter space that is still testable at the LHC.  In recent analyses from the ATLAS and CMS experiments it was found that  for some specific cases the lower mass bound for the chargino can go up to 600 GeV or even up to 1 TeV \cite{Sirunyan:2017lae, chargino-600}. We can see from Figure \ref{fig:msq-mx} that even for the case of 1 TeV  lower bound on the chargino mass, it is possible to have a squark sector that can be accessible at the LHC and, in addition, accommodate solutions for the muon $g-2$ anomaly (\textit{orange} points) which will be tested in the upcoming Fermilab experiment.

\begin{figure}
 \begin{subfigure}{.49\textwidth}
        \centering
        CMSSM
        \includegraphics[width=\linewidth]{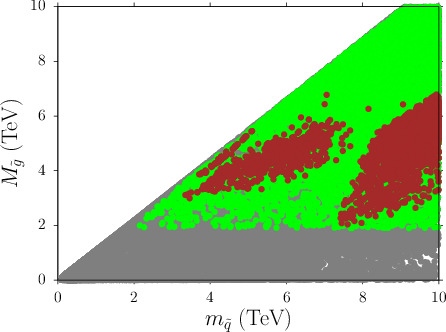}
    \end{subfigure}\hfill
\vspace{1cm}
    \begin{subfigure}{.49\textwidth}
        \centering
       NUHM2
        \includegraphics[width=\linewidth]{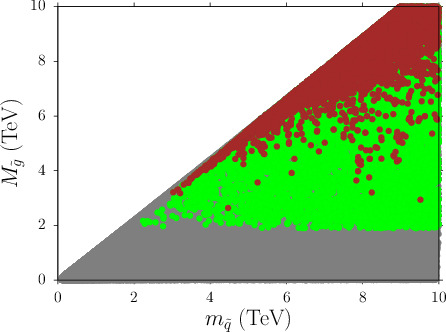}
    \end{subfigure}
\vspace{1cm}
        \begin{subfigure}{.49\textwidth}
        \centering
       SU(5)
        \includegraphics[width=\linewidth]{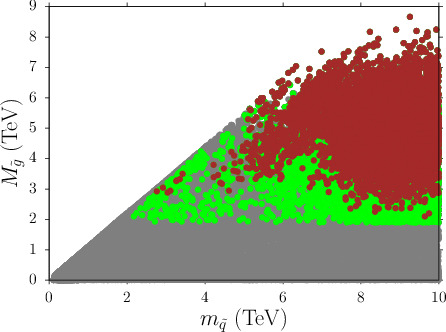}
    \end{subfigure}
        \begin{subfigure}{.49\textwidth}
        \centering
       SPLIT FAMILY NUHM2
        \includegraphics[width=\linewidth]{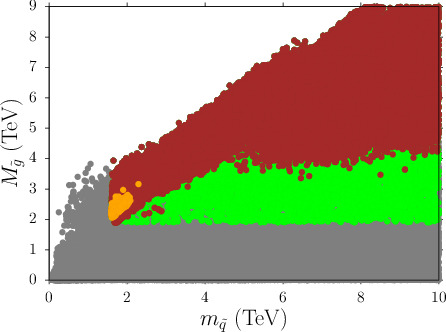}
    \end{subfigure}
    \begin{subfigure}{.49\textwidth}
        \centering
       NUGM+NUHM2
        \includegraphics[width=\linewidth]{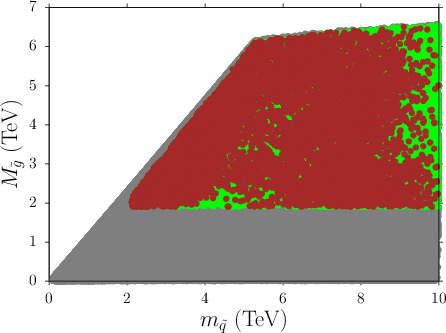}
    \end{subfigure}
        \begin{subfigure}{.49\textwidth}
        \centering
       SPLIT FAMILY NUGM
        \includegraphics[width=\linewidth]{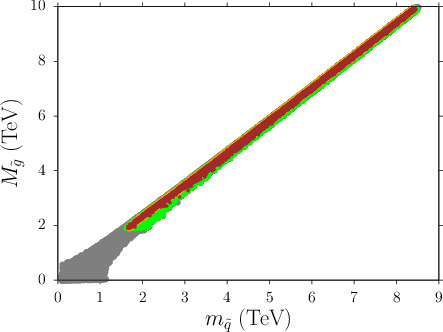}
    \end{subfigure}
\caption{Plots in the $m_{\tilde{g}}$ vs. $m_{\tilde{q}}$ plane. Color coding is the same as in Figure \ref{fig:mhf-m0}.}
\label{fig:mg-msq}
\end{figure}

Figure \ref{fig:mg-msq} displays our results in the $M_{\tilde{g}}$ vs. $m_{\tilde{q}}$ plane. The latest experimental searches for SUSY leads to a lower bound of 1.9 TeV on the gluino mass. In particular, the HL-LHC, with an anticipated luminosity of 3000 fb$^{-1}$, will be able to probe gluinos up to 2.3 TeV.  We can note from these figures that there is a considerable region of the parameter space that can still be tested at the current and future colliders. {Note that on the ``SPLIT FAMILY NUGM" panel there are plenty of solutions that satisfy the $g-2$ constraint (\textit{orange points}) under the brown points. Essentially, for the entire brown region in this panel there are points available to accommodate the muon $g-2$ current discrepancy.}

\begin{figure}
 \begin{subfigure}{.49\textwidth}
        \centering
        CMSSM
        \includegraphics[width=\linewidth]{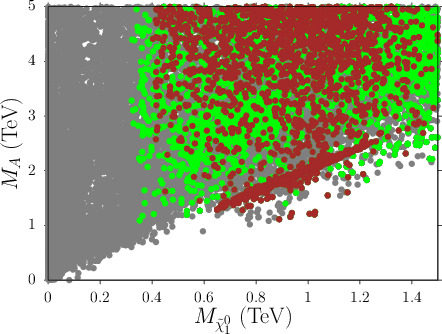}
    \end{subfigure}\hfill
\vspace{1cm}
    \begin{subfigure}{.49\textwidth}
        \centering
       NUHM2
        \includegraphics[width=\linewidth]{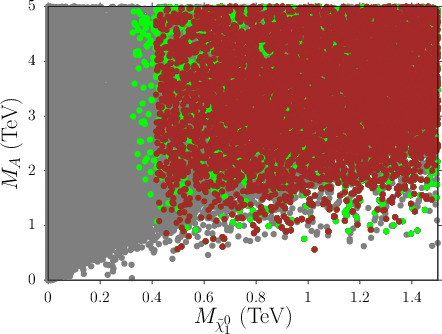}
    \end{subfigure}
\vspace{1cm}
        \begin{subfigure}{.49\textwidth}
        \centering
       SU(5)
        \includegraphics[width=\linewidth]{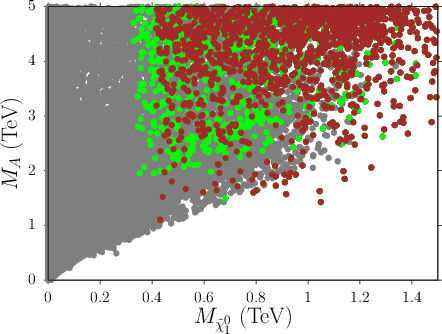}
    \end{subfigure}
        \begin{subfigure}{.49\textwidth}
        \centering
       SPLIT FAMILY NUHM2
        \includegraphics[width=\linewidth]{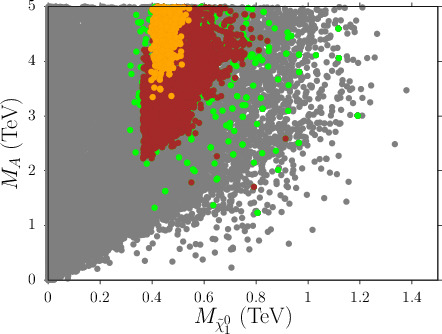}
    \end{subfigure}
    \begin{subfigure}{.49\textwidth}
        \centering
       NUGM+NUHM2
        \includegraphics[width=\linewidth]{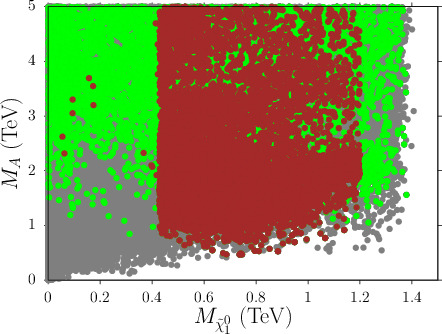}
    \end{subfigure}
        \begin{subfigure}{.49\textwidth}
        \centering
       SPLIT FAMILY NUGM
        \includegraphics[width=\linewidth]{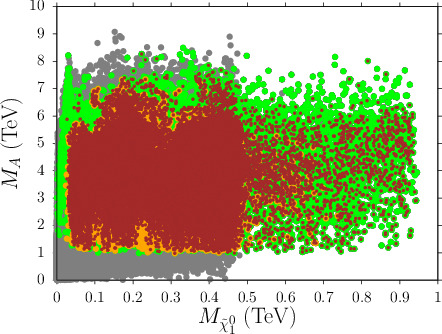}
    \end{subfigure}
\caption{Plots in the $M_A$ vs. $M_{\tilde{\chi}_1^0}$  plane. Color coding is the same as in Figure \ref{fig:mhf-m0}.}
\label{fig:ma-mx}
\end{figure}

In Figure \ref{fig:ma-mx} we display our results in the $M_A$ vs. $M_{\tilde{\chi}_1^0}$ and $M_{\tilde{\chi}_1^\pm}$ vs. $M_{\tilde{\chi}_1^0}$ planes. For the CMSSM we can note the presence of the A resonance channel $m_A \backsimeq 2 m_{\chi^0_1}$ to yield the desired relic abundance. There are, however, other coannihilation channels  that are clearly contributing in yielding the desired relic abundance. A detailed study of the nature of dark matter in these models will be performed in a follow up paper. We can also note from the figures that the pseudoscalar Higgs boson mass can be as light 500 GeV, which is within the reach of the HL-LHC.

In Figure \ref{fig:mch-msmu} we display our results in the $M_{\tilde{\chi}_1^\pm}$ vs. $m_{\tilde{\mu}_L}$ plane. The split family models allow for a light ${\tilde{\mu}_L}$ consistent with the desired relic abundance and $\gmu$. For the SF+NUHM2 model we find $m_{\tilde{\mu}_L}\gtrsim$ 600 GeV and for the SF+NUGM model it can be as light as $\sim$ 300 GeV. In addition, as described above, we can see from the \textit{orange} points in the figure that the smuons have to be light in order to make a sizable contribution to $\gmu$.

\begin{figure}
 \begin{subfigure}{.49\textwidth}
        \centering
        CMSSM
        \includegraphics[width=\linewidth]{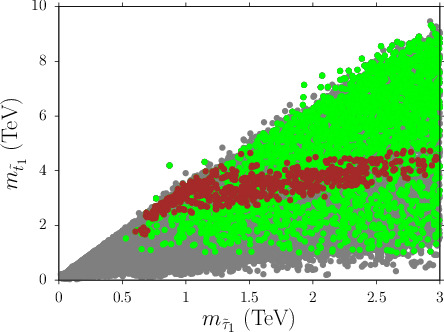}
    \end{subfigure}\hfill
\vspace{1cm}
    \begin{subfigure}{.49\textwidth}
        \centering
       NUHM2
        \includegraphics[width=\linewidth]{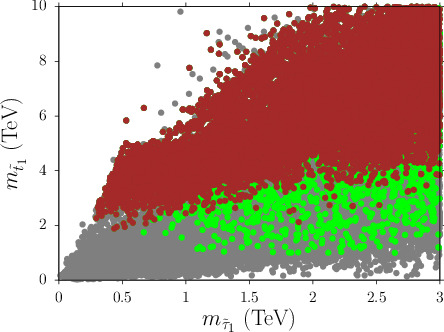}
    \end{subfigure}
\vspace{1cm}
        \begin{subfigure}{.49\textwidth}
        \centering
       SU(5)
        \includegraphics[width=\linewidth]{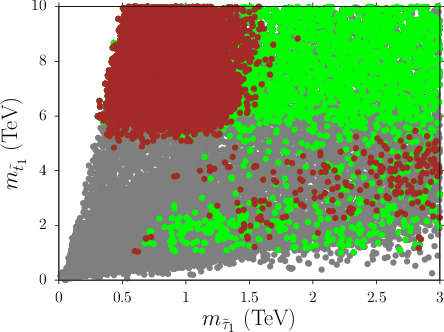}
    \end{subfigure}
        \begin{subfigure}{.49\textwidth}
        \centering
       SPLIT FAMILY NUHM2
        \includegraphics[width=\linewidth]{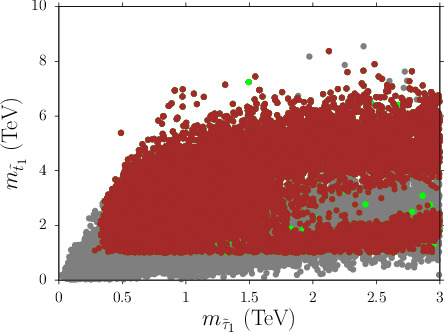}
    \end{subfigure}
    \begin{subfigure}{.49\textwidth}
        \centering
       NUGM+NUHM2
        \includegraphics[width=\linewidth]{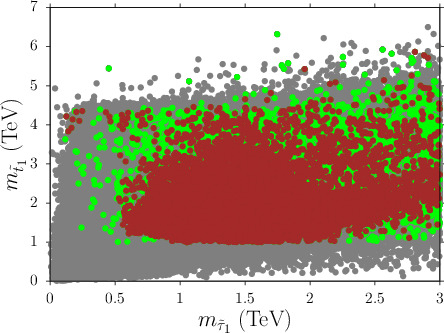}
    \end{subfigure}
        \begin{subfigure}{.49\textwidth}
        \centering
       SPLIT FAMILY NUGM
        \includegraphics[width=\linewidth]{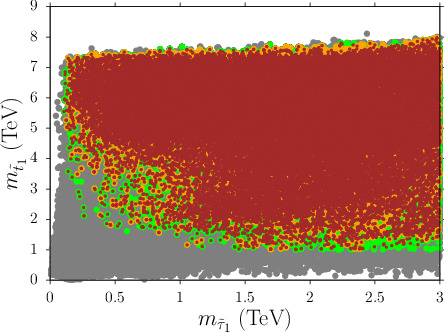}
    \end{subfigure}
\caption{Plots in the $m_{\tilde{t}_1}$ vs. $m_{\tilde{\tau}_1}$ plane. Color coding is the same as in Figure \ref{fig:mhf-m0}.}
\label{fig:mst-mstau}
\end{figure}

In Figure \ref{fig:mst-mstau} we display our results in the $m_{\tilde{t}_1}$ vs. $m_{\tilde{\tau}_1}$ plane. We can see that for each model the stau can be very light. In particular for the NUGM and SF+NUGM models the stau can be as light as 100 GeV. In addition, the light stau is also consistent with acceptable relic abundance and for the SF+NUGM model is also consistent with good $\gmu$.

\begin{figure}
 \begin{subfigure}{.49\textwidth}
        \centering
        CMSSM
        \includegraphics[width=\linewidth]{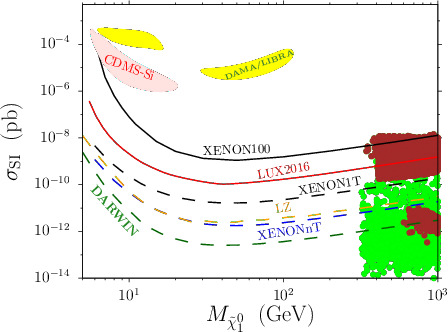}
    \end{subfigure}\hfill
\vspace{1cm}
    \begin{subfigure}{.49\textwidth}
        \centering
       NUHM2
        \includegraphics[width=\linewidth]{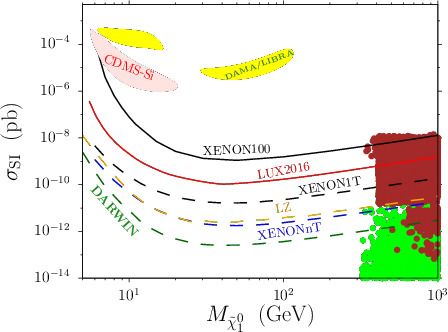}
    \end{subfigure}
\vspace{1cm}
        \begin{subfigure}{.49\textwidth}
        \centering
       SU(5)
        \includegraphics[width=\linewidth]{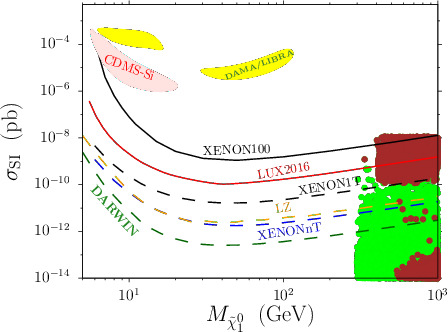}
    \end{subfigure}
        \begin{subfigure}{.49\textwidth}
        \centering
       SPLIT FAMILY NUHM2
        \includegraphics[width=\linewidth]{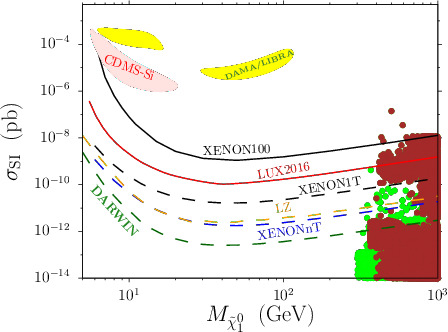}
    \end{subfigure}
    \begin{subfigure}{.49\textwidth}
        \centering
       NUGM+NUHM2
        \includegraphics[width=\linewidth]{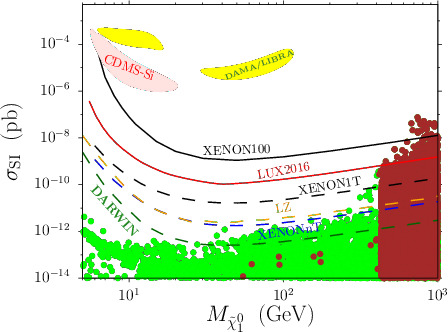}
    \end{subfigure}
        \begin{subfigure}{.49\textwidth}
        \centering
       SPLIT FAMILY NUGM
        \includegraphics[width=\linewidth]{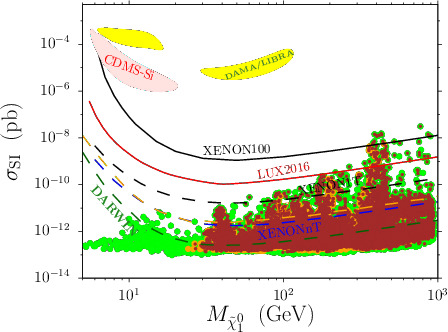}
    \end{subfigure}
\caption{Plots in the  $\sigma_{SI}$ vs. $M_{\tilde{\chi}_1^0}$ plane. Color coding is the same as in Figure \ref{fig:mhf-m0}.}
\label{fig:si-mx}
\end{figure}

In Figure \ref{fig:si-mx} we examine the prospects of direct detection of neutralino dark matter in our analysis in the $\sigma_{SI}$ vs. $M_{\tilde{\chi}_1^0}$ plane. The two anomalous signals DAMA/LIBRA \cite{Savage:2008er} and CDMS-Si \cite{Agnese:2013rvf} are shown in the upper left corner of the plot. In addition, we display the XENON100 \cite{Aprile:2016swn} and the LUX2016 bound~\cite{Akerib:2016vxi} (solid lines). The future projected reaches of XENON1T~\cite{Aprile:2015uzo}, LZ (with 1 keV cutoff)~\cite{Akerib:2015cja}, XENONnT~\cite{Aprile:2015uzo} and DARWIN ~\cite{Aalbers:2016jon} are shown as dashed lines. We can observe that the parameter space of all of these models can be probed by direct detection experiments as well. Except for the SF-NUGM model, a notable region of the parameter space of all the models is already excluded by the LUX experiment whereas a significant region is accessible to the XENON1T, LZ, XENONnT and DARWIN experiments. Most of the parameter space of the SF-NUGM model will be accessible to the XENON1T, LZ, XENONnT and DARWIN experiments. There is also a significant region of the parameter space of all the models (except SF-NUGM) with considerably low cross sections which is not accessible to any of the projected sensitivities.

\begin{figure}
 \begin{subfigure}{.49\textwidth}
        \centering
        CMSSM
        \includegraphics[width=\linewidth]{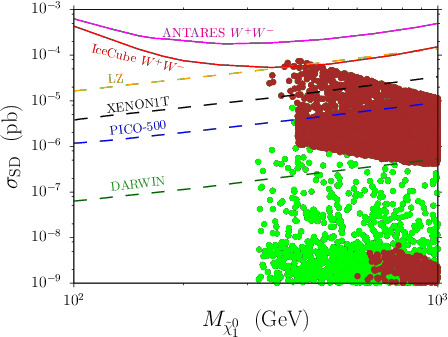}
    \end{subfigure}\hfill
\vspace{1cm}
    \begin{subfigure}{.49\textwidth}
        \centering
       NUHM2
        \includegraphics[width=\linewidth]{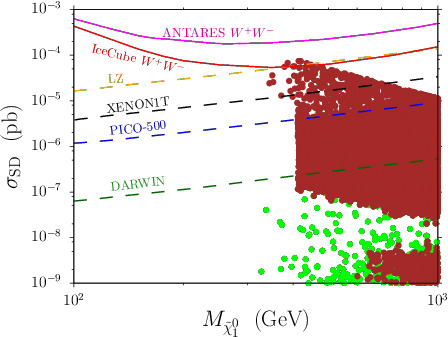}
    \end{subfigure}
\vspace{1cm}
        \begin{subfigure}{.49\textwidth}
        \centering
       SU(5)
        \includegraphics[width=\linewidth]{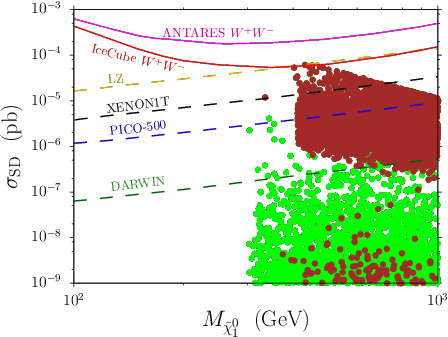}
    \end{subfigure}
        \begin{subfigure}{.49\textwidth}
        \centering
       SPLIT FAMILY NUHM2
        \includegraphics[width=\linewidth]{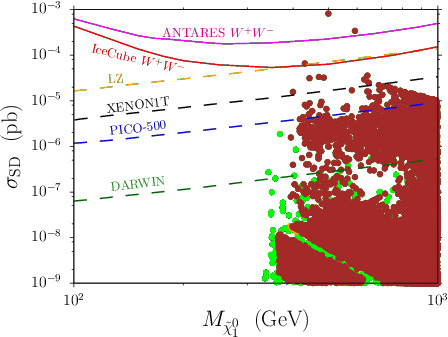}
    \end{subfigure}
    \begin{subfigure}{.49\textwidth}
        \centering
       NUGM+NUHM2
        \includegraphics[width=\linewidth]{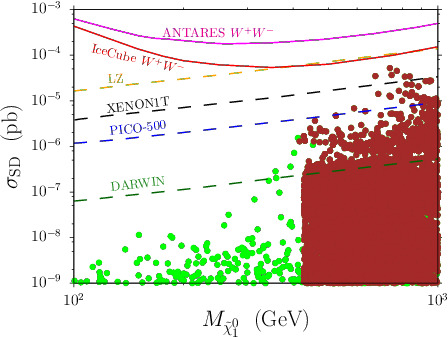}
    \end{subfigure}
        \begin{subfigure}{.49\textwidth}
        \centering
       SPLIT FAMILY NUGM
        \includegraphics[width=\linewidth]{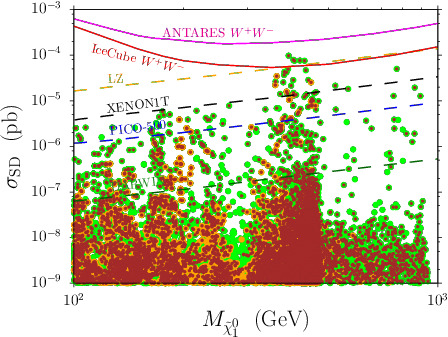}
    \end{subfigure}
\caption{Plots in the  $\sigma_{SD}$ vs. $M_{\tilde{\chi}_1^0}$ plane. Color coding is the same as in Figure \ref{fig:mhf-m0}.}
\label{fig:sd-mx}
\end{figure}

The spin dependent neutralino cross section is displayed in Figure \ref{fig:sd-mx} in the $\sigma_{SD}$ vs. $M_{\tilde{\chi}_1^0}$ plane. The recent limits from Antares ~\cite{Adrian-Martinez:2016gti} and IceCube~\cite{Aartsen:2016exj} are shown as solid lines. We can see that the parameter space of all the models is barely accessible to these experiments. The dashed lines show the projected reach of the LZ~\cite{Akerib:2015cja}, XENON1T~\cite{Aalbers:2016jon}, Pico-500~\cite{ckrauss} and the
 DARWIN~\cite{Aalbers:2016jon} experiments. We can see that the parameter space of all of these models is accessible to these future experiments. There is, however, a significant region of the parameter space of all the models, with low cross section, which is well beyond the search limit of all of these experiments.


\section{Conclusion}\label{sec:conclude}

We presented a status update on various models of supersymmetry in light of latest constraints from the LHC and direct detection experiments. We showed that a considerable region of the parameter space of SUSY is still accessible to these experiments and continued experimental search is crucial.  We also showed that a notable region of the parameter space of all of the analyzed models yields the desired relic abundance and is accessible to future direct detection experiments. Furthermore, we found solutions in the parameter space of the split family models that are  consistent with the $\gmu$ constraint.

\section{Acknowledgments}

Work of IG is supported in part by Bartol Research Institute.


\newpage


\begin{figure}
 \begin{subfigure}{.49\textwidth}
        \centering
        CMSSM
        \includegraphics[width=\linewidth]{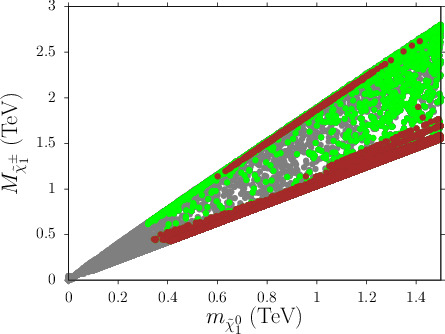}
    \end{subfigure}\hfill
\vspace{1cm}
    \begin{subfigure}{.49\textwidth}
        \centering
       NUHM2
        \includegraphics[width=\linewidth]{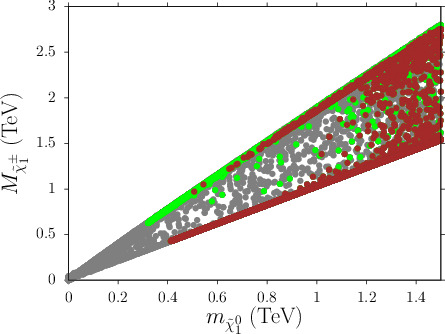}
    \end{subfigure}
\vspace{1cm}
        \begin{subfigure}{.49\textwidth}
        \centering
       SU(5)
        \includegraphics[width=\linewidth]{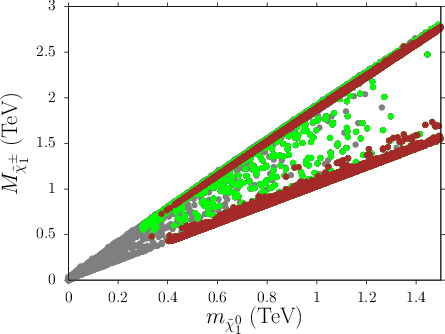}
    \end{subfigure}
        \begin{subfigure}{.49\textwidth}
        \centering
       SPLIT FAMILY NUHM2
        \includegraphics[width=\linewidth]{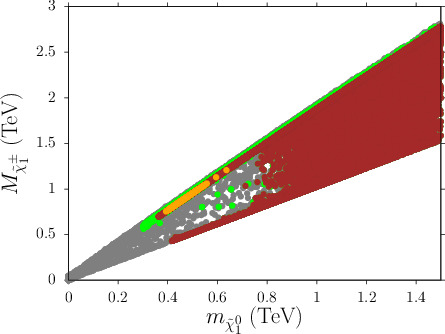}
    \end{subfigure}
    \begin{subfigure}{.49\textwidth}
        \centering
       NUGM+NUHM2
        \includegraphics[width=\linewidth]{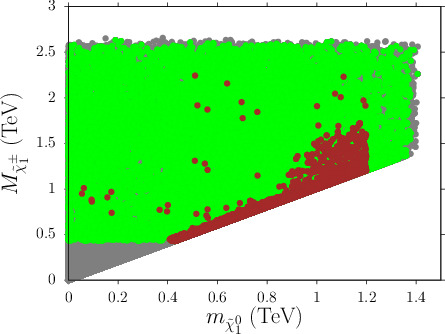}
    \end{subfigure}
        \begin{subfigure}{.49\textwidth}
        \centering
       SPLIT FAMILY NUGM
        \includegraphics[width=\linewidth]{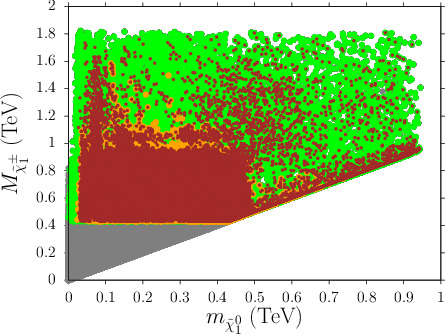}
    \end{subfigure}
\caption{Plots in the $M_{\tilde{\chi}_1^\pm}$ vs. $m_{\tilde{\chi}_1^0}$  plane. Color coding is the same as in Figure \ref{fig:mhf-m0}.}
\label{fig:mch-mx}
\end{figure}



\begin{thebibliography}{99}

\bibitem{Jungman:1995df}
  See, for instance, G.~Jungman, M.~Kamionkowski and K.~Griest,
  Phys.\ Rept.\  {\bf 267}, 195 (1996).


%
\bibitem{lhc-squark} ATLAS collaboration, ATLAS-CONF-2017-022; CMS
  Collaboration, CMS-SUS-16-036.
%
\bibitem{lhc-wino} ATLAS Collaboration, ATLAS-CONF-2017-039; CMS
  Collaboration, CMS-16-034.
%
\bibitem{lhc-stop} ATLAS Collboration, ATLAS-CONF-2017-020; CMS
  Collboration, CMS-SUS-16-051 and CMS-SUS-16-049.

\bibitem{lhc-chargino} ATLAS Collboration, ATLAS-CONF-2017-017.
%
\bibitem{upgrade} See, {\it e.g.} ATLAS Phys. PUB 2013-011; CMS
  Note-13-002.
%
\bibitem{gershtein}
Y.~Gershtein {\it et al.},
arXiv:1311.0299 [hep-ex].

%
\bibitem{atlaswiki}
\verb^https://twiki/cern.ch/bin/view/AtlasPublic/UpgradePhysicsStudies^

\bibitem{Baer:2017pba}
  See, for instance,
  A.~Aboubrahim and P.~Nath,
  Phys.\ Rev.\ D {\bf 96}, 075015 (2017);
  H.~Baer, V.~Barger, J.~S.~Gainer, H.~Serce and X.~Tata,
  arXiv:1708.09054 [hep-ph];
  K.~Kowalska, L.~Roszkowski, E.~M.~Sessolo and A.~J.~Williams,
  JHEP {\bf 1506}, 020 (2015);
  C.~Han, K.~i.~Hikasa, L.~Wu, J.~M.~Yang and Y.~Zhang,
  Phys.\ Lett.\ B {\bf 769}, 470 (2017);
  W.~Ahmed, X.~J.~Bi, T.~Li, J.~S.~Niu, S.~Raza, Q.~F.~Xiang and P.~F.~Yin,
  arXiv:1709.06371 [hep-ph];
  J.~Chakrabortty, A.~Choudhury and S.~Mondal,
  JHEP {\bf 1507}, 038 (2015);
  J.~Kawamura and Y.~Omura,
  JHEP {\bf 1708}, 072 (2017);
  J.~Kawamura and Y.~Omura,
  Phys.\ Rev.\ D {\bf 93}, no. 5, 055019 (2016).



\bibitem{Hagiwara:2011af}
  M.~Davier, A.~Hoecker, B.~Malaescu and Z.~Zhang,
  Eur.\ Phys.\ J.\ C {\bf 71}, 1515 (2011)
  [Erratum-ibid.\ C {\bf 72}, 1874 (2012)];
  K.~Hagiwara, R.~Liao, A.~D.~Martin, D.~Nomura and T.~Teubner,
  J.\ Phys.\ G {\bf 38}, 085003 (2011).


\bibitem{Ajaib:2014ana}
  M.~A.~Ajaib, I.~Gogoladze, Q.~Shafi and C.~S.~Un,
  JHEP {\bf 1405}, 079 (2014).

\bibitem{Moroi:1995yh}
  T.~Moroi,
  Phys.\ Rev.\ D {\bf 53}, 6565 (1996)
  [Erratum-ibid.\ D {\bf 56}, 4424 (1997)];


\bibitem{Martin:2001st}
  S.~P.~Martin and J.~D.~Wells,
  Phys.\ Rev.\ D {\bf 64}, 035003 (2001);
  G.~F.~Giudice, P.~Paradisi and A.~Strumia,
  JHEP {\bf 1210}, 186 (2012).

\bibitem{ISAJET}
  F.~E.~Paige, S.~D.~Protopopescu, H.~Baer and X.~Tata,
  hep-ph/0312045.

\bibitem{Belanger:2008sj}
  G.~Belanger, F.~Boudjema, A.~Pukhov and A.~Semenov,
  Comput.\ Phys.\ Commun.\  {\bf 180}, 747 (2009).


\bibitem{Leva}
J.L. Leva,
 Math. Softw. 18 (1992) 449;
J.L. Leva,
Math. Softw. 18 (1992) 454.



\bibitem{Ajaib:2015ika}
  M.~Adeel Ajaib, I.~Gogoladze and Q.~Shafi,
  Phys.\ Rev.\ D {\bf 91}, no. 9, 095005 (2015).


\bibitem{Kane:1993td}
  G.~L.~Kane, C.~F.~Kolda, L.~Roszkowski and J.~D.~Wells,
  Phys.\ Rev.\ D {\bf 49}, 6173 (1994);
  J.~Ellis, J.~L.~Evans, A.~Mustafayev, N.~Nagata and K.~A.~Olive,
  Eur.\ Phys.\ J.\ C {\bf 76}, no. 11, 592 (2016) and references therein.



\bibitem{Chamseddine:1982jx}
A.~H.~Chamseddine, R.~L.~Arnowitt and P.~Nath,
  Phys.\ Rev.\ Lett.\  {\bf 49}, 970 (1982);
 R.~Barbieri, S.~Ferrara and C.~A.~Savoy,
  Phys.\ Lett.\ B {\bf 119}, 343 (1982);
     L.~J.~Hall, J.~D.~Lykken and S.~Weinberg,
  Phys.\ Rev.\ D {\bf 27}, 2359 (1983);
  N. Ohta,
Prog. Theor. Phys. 70 (1983) 542.


\bibitem{Profumo:2003ema}
  S.~Profumo,
  Phys.\ Rev.\  D {\bf 68}, 015006 (2003);
  B.~Ananthanarayan and P.~N.~Pandita,
  Int.\ J.\ Mod.\ Phys.\  A {\bf 22}, 3229 (2007).

\bibitem{Gogoladze:2008dk}
  I.~Gogoladze, R.~Khalid, N.~Okada and Q.~Shafi,
  Phys.\ Rev.\ D {\bf 79}, 095022 (2009);
  H.~Baer, I.~Gogoladze, A.~Mustafayev, S.~Raza and Q.~Shafi,
  JHEP {\bf 1203}, 047 (2012).



\bibitem{Ellis:2002wv}
  J.~R.~Ellis, K.~A.~Olive and Y.~Santoso,
  Phys.\ Lett.\ B {\bf 539}, 107 (2002);
  J.~R.~Ellis, T.~Falk, K.~A.~Olive and Y.~Santoso,
  Nucl.\ Phys.\ B {\bf 652}, 259 (2003);
  H.~Baer, A.~Mustafayev, S.~Profumo, A.~Belyaev and X.~Tata,
  JHEP {\bf 0507}, 065 (2005).


\bibitem{Babu:2014sga}
  K.~S.~Babu, I.~Gogoladze, S.~Raza and Q.~Shafi,
  Phys.\ Rev.\ D {\bf 90}, no. 5, 056001 (2014).


\bibitem{Martin:2009ad}
 B.~Ananthanarayan, P.~N.~Pandita,
  Int.\ J.\ Mod.\ Phys.\  {\bf A22}, 3229-3259 (2007);
  S.~Bhattacharya, A.~Datta and B.~Mukhopadhyaya,
  JHEP {\bf 0710}, 080 (2007);
   S.~P.~Martin,
  Phys.\ Rev.\  {\bf D79}, 095019 (2009);
  J.~Chakrabortty and A.~Raychaudhuri,
  Phys.\ Lett.\ B {\bf 673}, 57 (2009);
  J.~Chakrabortty, S.~Mohanty and S.~Rao,
  JHEP {\bf 1402}, 074 (2014).


\bibitem{Martin:2013aha}
  S.~P.~Martin,
  Phys.\ Rev.\ D {\bf 89}, no. 3, 035011 (2014).


\bibitem{Anandakrishnan:2013cwa}
  A.~Anandakrishnan and S.~Raby,
  Phys.\ Rev.\ Lett.\  {\bf 111}, 211801 (2013);
  B.~Zhu, R.~Ding and T.~Li,
  arXiv:1701.05511 [hep-ph].

\bibitem{Babu:2014lwa}
  K.~S.~Babu, I.~Gogoladze, Q.~Shafi and C.~S.~Un,
  Phys.\ Rev.\ D {\bf 90}, no. 11, 116002 (2014).

\bibitem{Gogoladze:2012yf}
 See, for instance,  I.~Gogoladze, F.~Nasir and Q.~Shafi,
  Int.\ J.\ Mod.\ Phys.\ A {\bf 28}, 1350046 (2013):
  I.~Gogoladze, F.~Nasir and Q.~Shafi,
  JHEP {\bf 1311}, 173 (2013);
  Z.~Kang, J.~Li and T.~Li,
  JHEP {\bf 1211}, 024 (2012);
  A.~Cici, Z.~Kirca and C.~S.~Un,
  arXiv:1611.05270 [hep-ph].

\bibitem{Baer:2004xx}
  H.~Baer, A.~Belyaev, T.~Krupovnickas and A.~Mustafayev,
  JHEP {\bf 0406}, 044 (2004).



\bibitem{:2009ec}
    [Tevatron Electroweak Working Group and CDF Collaboration and D0 Collab],
  arXiv:0903.2503 [hep-ex].





\bibitem{Pokoroski}
  M.~Badziak, Z.~Lalak, M.~Lewicki, M.~Olechowski and S.~Pokorski,
  JHEP {\bf 1503}, 003 (2015);
  F.~Wang, W.~Wang and J.~M.~Yang,
  JHEP {\bf 1506}, 079 (2015).

\bibitem{Sirunyan:2017lae}
  A.~M.~Sirunyan {\it et al.} [CMS Collaboration],
  arXiv:1709.05406 [hep-ex].
  
\bibitem{chargino-600} ATLAS Collaboration, ATLAS-CONF-2017-039; CMS Collaboration, CMS-16-034.

\bibitem{:2009ec}
    [Tevatron Electroweak Working Group and CDF Collaboration and D0 Collab],
  arXiv:0903.2503 [hep-ex].





\bibitem{Pokoroski}
  M.~Badziak, Z.~Lalak, M.~Lewicki, M.~Olechowski and S.~Pokorski,
  JHEP {\bf 1503}, 003 (2015)
  [arXiv:1411.1450 [hep-ph]].



%
\bibitem{Aprile:2016swn}
  E.~Aprile {\it et al.} [XENON100 Collaboration],
  arXiv:1609.06154 [astro-ph.CO].
  
\bibitem{Savage:2008er} 
  C.~Savage, G.~Gelmini, P.~Gondolo and K.~Freese,
  JCAP {\bf 0904}, 010 (2009)
  doi:10.1088/1475-7516/2009/04/010
  [arXiv:0808.3607 [astro-ph]].
  
\bibitem{Agnese:2013rvf} 
  R.~Agnese {\it et al.} [CDMS Collaboration],
  Phys.\ Rev.\ Lett.\  {\bf 111}, no. 25, 251301 (2013)
  doi:10.1103/PhysRevLett.111.251301
  [arXiv:1304.4279 [hep-ex]].
  
\bibitem{Akerib:2016vxi}
  D.~S.~Akerib {\it et al.},
  arXiv:1608.07648 [astro-ph.CO].
  
%
\bibitem{Akerib:2015cja}
  D.~S.~Akerib {\it et al.} [LZ Collaboration],
  arXiv:1509.02910 [physics.ins-det].


%
\bibitem{Aprile:2015uzo}
  E.~Aprile {\it et al.} [XENON Collaboration],
  JCAP {\bf 1604} (2016) no.04,  027
  doi:10.1088/1475-7516/2016/04/027
  [arXiv:1512.07501 [physics.ins-det]].



%
\bibitem{Aalbers:2016jon}
  J.~Aalbers {\it et al.} [DARWIN Collaboration],
  arXiv:1606.07001 [astro-ph.IM].
%

%
\bibitem{Adrian-Martinez:2016gti}
  S.~Adrian-Martinez {\it et al.} [ANTARES Collaboration],
  Phys.\ Lett.\ B {\bf 759} (2016) 69
  doi:10.1016/j.physletb.2016.05.019
  [arXiv:1603.02228 [astro-ph.HE]].
%

%
\bibitem{Aartsen:2016exj}
  M.~G.~Aartsen {\it et al.} [IceCube Collaboration],
  JCAP {\bf 1604} (2016) no.04,  022
  doi:10.1088/1475-7516/2016/04/022
  [arXiv:1601.00653 [hep-ph]].
%
\bibitem{ckrauss} Talk by C. Krauss for the Pico collaboration, ICHEP 2016 meeting, Chicago, IL, August 2016.


\end{thebibliography}
\end{document}